\documentstyle[fleqn,12pt]{article}
\catcode`@=11

\font\twlmsx=msam10 scaled 1200
\font\egtmsx=msam7 scaled 1200
\font\fivemsx=msam5

\font\twlmsy=msbm10 scaled 1200
\font\egtmsy=msbm7 scaled 1200
\font\fivemsy=msbm5

\font\twleufm=eufm12
\font\egteufm=eufm7 scaled 1200
\font\fiveufm=eufm5

\newfam\msxfam
\newfam\msyfam
\newfam\eufmfam

\textfont\msxfam=\twlmsx  \scriptfont\msxfam=\egtmsx
\scriptscriptfont\msxfam=\fivemsx

\textfont\msyfam=\twlmsy  \scriptfont\msyfam=\egtmsy
\scriptscriptfont\msyfam=\fivemsy

\textfont\eufmfam=\twleufm \scriptfont\eufmfam=\egteufm
\scriptscriptfont\eufmfam=\fiveufm

\def\hexnumber@#1{\ifcase#1 0\or1\or2\or3\or4\or5\or6\or7\or8\or9\or
	A\or B\or C\or D\or E\or F\fi }
\edef\msx@{\hexnumber@\msxfam}
\edef\msy@{\hexnumber@\msyfam}
\edef\eufm@{\hexnumber@\eufmfam}
\def\Bbb{\ifmmode\let\next\Bbb@\else
 \def\next{\errmessage{Use \string\Bbb\space only in math mode}}\fi\next}
\def\Bbb@#1{{\Bbb@@{#1}}}
\def\Bbb@@#1{\fam\msyfam#1}

\mathchardef\AA="0\msy@41
\mathchardef\BB="0\msy@42
\mathchardef\CC="0\msy@43
\mathchardef\DD="0\msy@44
\mathchardef\EE="0\msy@45
\mathchardef\FF="0\msy@46
\mathchardef\GG="0\msy@47
\mathchardef\HH="0\msy@48
\mathchardef\II="0\msy@49
\mathchardef\JJ="0\msy@4A
\mathchardef\KK="0\msy@4B
\mathchardef\LL="0\msy@4C
\mathchardef\MM="0\msy@4D
\mathchardef\NN="0\msy@4E
\mathchardef\OO="0\msy@4F
\mathchardef\PP="0\msy@50
\mathchardef\QQ="0\msy@51
\mathchardef\RR="0\msy@52
\mathchardef\SS="0\msy@53
\mathchardef\TT="0\msy@54
\mathchardef\UU="0\msy@55
\mathchardef\VV="0\msy@56
\mathchardef\WW="0\msy@57
\mathchardef\XX="0\msy@58
\mathchardef\YY="0\msy@59
\mathchardef\ZZ="0\msy@5A

\mathchardef\frA="0\eufm@41
\mathchardef\frB="0\eufm@42
\mathchardef\frC="0\eufm@43
\mathchardef\frD="0\eufm@44
\mathchardef\frE="0\eufm@45
\mathchardef\frF="0\eufm@46
\mathchardef\frG="0\eufm@47
\mathchardef\frH="0\eufm@48
\mathchardef\frI="0\eufm@49
\mathchardef\frJ="0\eufm@4A
\mathchardef\frK="0\eufm@4B
\mathchardef\frL="0\eufm@4C
\mathchardef\frM="0\eufm@4D
\mathchardef\frN="0\eufm@4E
\mathchardef\frO="0\eufm@4F
\mathchardef\frP="0\eufm@50
\mathchardef\frQ="0\eufm@51
\mathchardef\frR="0\eufm@52
\mathchardef\frS="0\eufm@53
\mathchardef\frT="0\eufm@54
\mathchardef\frU="0\eufm@55
\mathchardef\frV="0\eufm@56
\mathchardef\frW="0\eufm@57
\mathchardef\frX="0\eufm@58
\mathchardef\frY="0\eufm@59
\mathchardef\frZ="0\eufm@5A

\mathchardef\fra="0\eufm@61
\mathchardef\frb="0\eufm@62
\mathchardef\frc="0\eufm@63
\mathchardef\frd="0\eufm@64
\mathchardef\fre="0\eufm@65
\mathchardef\frf="0\eufm@66
\mathchardef\frg="0\eufm@67
\mathchardef\frh="0\eufm@68
\mathchardef\fri="0\eufm@69
\mathchardef\frj="0\eufm@6A
\mathchardef\frk="0\eufm@6B
\mathchardef\frl="0\eufm@6C
\mathchardef\frm="0\eufm@6D
\mathchardef\frn="0\eufm@6E
\mathchardef\fro="0\eufm@6F
\mathchardef\frp="0\eufm@70
\mathchardef\frq="0\eufm@71
\mathchardef\frr="0\eufm@72
\mathchardef\frs="0\eufm@73
\mathchardef\frt="0\eufm@74
\mathchardef\fru="0\eufm@75
\mathchardef\frv="0\eufm@76
\mathchardef\frw="0\eufm@77
\mathchardef\frx="0\eufm@78
\mathchardef\fry="0\eufm@79
\mathchardef\frz="0\eufm@7A
\catcode`@=11

\newcommand{\bea}{\begin{eqnarray*}}
\newcommand{\eea}{\end{eqnarray*}}

\newcommand{\ben}{\begin{equation}} \newcommand{\een}{\end{equation}}

\newtheorem{theorem}{Theorem}
\newtheorem{lemma}[theorem]{Lemma}
\begin{document}
\thispagestyle{empty}
\vspace*{\fill}
%\vspace*{55mm}
\centerline{\Large\bf Vertex Operators are not closeable}
\vspace{10mm}
\begin{center}{\large Wolfram Boenkost}\\
\vspace{5mm}
Fachbereich Mathematik\\
J. W. Goethe-Universitt Frankfurt\\
Postfach 11 19 32\\
D-60054 Frankfurt a. M.\\
Germany \\
\end{center}
\vfill
\centerline{\today}
\vfill
\noindent {\bf Abstract:}\\ We prove the uncloseability of the free vertex
operators used in conformal field theory for the BRST--construction of
primary fields. Our proof includes minimal models as well as WZNW--models.
\vfill
\newpage\setcounter{page}{1}
\section{Introduction}
Free vertex operators in Fock space have proved to be very useful in the
context of conformal quantum field theory in two dimensions. Beginning
with the work of of A. Tsuchiya, Y. Kanie (\cite{TK1}, \cite{TK2}) and G.
Felder \cite{Fe}, who constructed the primary fields of the minimal models
using vertex operators, several authors (\cite{BF}, \cite{Ku}, \cite{BMP})
generalized this construction to WZNW models. The question arises, wether
there is an operator content in these constructions, i.e. if the fields
constructed fulfill the axioms of some axiomatic field theory. In this
note we want to prove, continuing the work in \cite{BC}, a negative result
saying that the building blocks in these constructions, the vertex
operators, are noncloseable operators in Fock space. The main reason for
this is the fact that the complex chiral decomposition inherent in
relating vertex operators to conformal two dimensional fields is not
compatible with a sound Hilbert space interpretation of free vertices.
Certainly a pure Minkowski space approach to the operator content of the
two dimensional conformal field theory is not excluded, but the analytic
continuation seems to produce unusual (and unexpected) objects.

We restrict ourself to the vertices used for minimal models, to avoid the
expressions to be too nasty. Nevertheless  due to the similarity of the
vertex operators for minimal models and vertex operators for WZNW models
in essential points, our proof will also work in these cases. We will
discuss this point later on. It should be noted that for WZNW models the
Fock space is no longer a Hilbert space, but a Krein space. But the
question of closeability depends not on this point \cite{Bo}.
\section{Statement of the result}
Let $\left\{ a_{\pm n} : n \in \NN\right\}$ be the generators of a
Heisenberg algebra with commutation relations $[a_n,a_m]=n\delta_{n,-m}$
represented on the Fock space $\cal H$ with scalar product
$\langle\cdot|\cdot\rangle$, vacuum state $\Phi_0$ and canonical basis
$\{\Phi_{\alpha}\}$  indexed by the multiindices
$\alpha=(\alpha_1,\ldots)$ fulfilling $\alpha_i \ge 0$ and $\|\alpha\| :=
\sum i \alpha_i < \infty$. Vertex operators are defined by
\ben\label{defi}
V(\gamma,z)=\exp\left(\gamma \sum_{n=1}^{\infty} \frac{z^n}{n}
a_{-n}\right) \exp\left(-\gamma \sum_{n=1}^{\infty} \frac{z^{-n}}{n} a_n
\right)
\een
for $\gamma,z \in \CC, z \neq 0$. In \cite{BC} we have proved that
(\ref{defi}) defines a densely defined operator in $\cal H$ for $|z|<1$.
There we have defined $V(\gamma,z)$ as the maximal matrix operator in
$\cal H$, defined by matrix elements $\langle
\Phi_{\alpha}|V(\gamma,z)\Phi_{\beta}\rangle$, which can be easily
calculated.

Since the closeability of an operator may depend on its domain of
definition, we introduce also the minimal operator $V(\gamma,z)_0$ with
domain of definition $D(V(\gamma,z)_0)=D=\mbox{Lin}\{\Phi_{\alpha}\}$. We
have $V(\gamma,z)_0\subset V(\gamma,z)$. We shall prove in this note the
following theorem.
\begin{theorem} Let $V(\gamma,z)$ and $V(\gamma,z)_0$ as above. Then
\begin{itemize}
\item[(i)] If $|z|<1$, then $V(\gamma,z)$ and $V(\gamma,z)_0$ are
noncloseable in $\cal H$.
\item[(ii)] If $|z|>1$, then the domain of definition of $V(\gamma,z)$ is
$\{0\}$ (in this case $V(\gamma,z)$ is trivially closed).
\end{itemize}
\end{theorem}
Note that obviously, for $|z|<1$, any extension of $V(\gamma,z)_0$ is also
noncloseable.\\
The cases (i) and (ii) are equivalent by the relation
$V(\gamma,z)_0^*=V(-\bar{\gamma},1/\bar{z})$ (cf. \cite{Wd}, Thm. 6.20).
\section{Coherent states}
The additional input we need are the coherent states, which give
eigenstates for the annihilation operators and are defined as follows. Let
$l\in \NN$, $\beta \in \CC$. Then for $t=(t_1,\ldots,t_l)\in \CC^l$ let
\ben\label{co1}
|t,\beta,l\rangle := N_l \sum_{\alpha_1,\ldots,\alpha_l=0}^{\infty}
\prod_{i=1}^l \frac{(\beta t_i)^{\alpha_i}}{(\alpha_i!
i^{\alpha_i})^{1/2}} \Phi_{\alpha}
\een
with $N_l=\exp\left(-\frac{|\beta|^2}{2} \sum_{n=0}^l
\frac{|t_n|^2}{n}\right)$. The states $|t,\beta,l\rangle$ are called
coherent states and have the properties $\|\; |t,\beta,l\rangle\|=1$,
$|t,\beta,l\rangle\in D(a_{\pm n})$ and
\ben
a_n|t,\beta,l \rangle = \cases{ \beta t_n | t,\beta,l\rangle & for $0<n\le
l$, \cr
0 & for $n>l$.\cr}
\een
Let $l_{2,h}$ be the space of complex sequences $t=(t_1,t_2,\ldots)$ with
$\sum |t_i|^2/i < \infty$. For $t\in l_{2,h}$ the limit $\lim_{l\to\infty}
|t,\beta,l\rangle =: |t,\beta\rangle$ exists in $\cal H$. We have again
$|t,\beta\rangle\in D(a_{\pm n})$ and $a_n|t,\beta\rangle=\beta t_n |
t,\beta\rangle$ for $n\ge 0$. We can write (\ref{co1}) in an equivalent
form
\ben\label{co2}
|t,\beta,l\rangle=N_l \exp \left(\beta \sum_{n=1}^{l} \frac{t_n}{n} a_n
\right) \Phi_0
\een
where the exponential is defined by its power series. From (\ref{co2}) we
see, that up to the normalization constant $N_l$, $|t,\beta,l\rangle$ is
given by the action of a generalized vertex operator as introduced in
\cite{BC} on the ground state $\Phi_0$.

The coherent states form an overcomplete set in $\cal H$, we want to make
use of this statement in the following form. Let $l_{2,0}$ be the set of
finite sequences.
\begin{lemma} For any $t\in l_{2,h}$ the set
$\big\{ |t+t',\beta\rangle,\;t'\in l_{2,0} \big\}
$ is total in $\cal H$.
\end{lemma}
{\bf Remark:} Results on (over--) completeness of coherent states are well
known. For one degree of freedom (i.e. one pair of creation and
annihilation operators) much stronger results are known (cf. Bargmann e.a.
\cite{Ba}). For a finite number of degrees of freedom the arguments can be
easily adapted from the one dimensional case. But for our case with a
countable number of degrees of freedom we do not know rigouros results of
the type stated here. We sketch a proof of Lemma 2 in the appendix.
\section{Proof of Theorem 1}
We prove first that  the maximal operator is noncloseable for $|z|<1$. The
idea is to look at $\langle \Phi|V(\gamma,z)f\rangle$ for any $\Phi \in
{\cal H}$ and $f\in D(V(\gamma,z))$. If $\Phi\in D(V(\gamma,z)^*)$ the
mapping $f\mapsto \langle \Phi|V(\gamma,z) f\rangle$ must be continous on
$D(V(\gamma,z))$. But for any $\Phi \neq 0$ we shall construct a sequence
of coherent states $t_n$ in $D(V(\gamma,z))$ for which $|\langle
\Phi|V(\gamma,z)t_n\rangle | \to \infty$ for $n\to\infty$.

In \cite{BC} we have proved a composition theorem for vertex operators and
proved that the well known product formula for vertex operators holds  on
$D$ in the strong sense. Since coherent states can be viewed as
(generalized) vertex operators we can apply this formula and get
immediately the following Lemma.
\begin{lemma}Let $0<|z|<1$. Then
\begin{itemize}
\item[(i)] $|t,\beta,l\rangle \in D(V(\gamma,z)).$
\item[(ii)] $V(\gamma,z) |t,\beta,l\rangle = K^{-1}\exp(-\gamma\beta
\sum\limits_{n=1}^l {\displaystyle\frac{z^{-n}t_n}{n}}) |\tau,l\rangle,$
where $|\tau,l\rangle$ is  a coherent state, we have
\ben
a_n |\tau,l\rangle=\cases{(\gamma z^n + \beta t_n) | \tau,l\rangle & for
$0<n\le l$,\cr
\gamma z^n|\tau,l\rangle& for $n>l$.}
\een
\end{itemize}
The constant is given by $K=\exp(\displaystyle\frac{1}{2}
\sum_{n=1}^{\infty}
\frac{|\gamma z^n+\beta t_n|^2 - |\beta t_n|^2}{n} )$.
\end{lemma}
Now let $\gamma,z\in \CC$ with $0< |z|<1$ be fixed. Choose a sequence
$t\in l_{2,h}$ and $\beta \in \CC$ with
\ben
\lim_{l\to \infty} | \exp (-\gamma \beta \sum_{n=1}^l \frac{z^{-n} t_n}{n}
)| = \infty.
\een
For example we can choose $t_n=1/n$ and $\beta=-\bar{\gamma}$. Let
$\Phi\in{\cal H}, \Phi \ne 0$. Then by Lemma 2, there exists $t'\in
l_{2,0}$ with $\langle \Phi|\tau + t'\rangle \ne 0$, where
$|\tau\rangle=\lim_{l\to\infty}|\tau,l\rangle$. \\
Since $\langle \Phi|\tau+t',l\rangle \longrightarrow \langle
\Phi|\tau+t'\rangle$ ($l\to \infty$) we have for $l$ big enough
$0<c<|\langle \Phi|\tau+t',l\rangle|$.
Hence we get
{\mathindent5mm\ben
|\langle \Phi|V(\gamma,z)|t+t',\beta,l\rangle|=|\exp (-\gamma \beta
\sum_{n=1}^l \frac{z^{-n} (t_n+t'_n)}{n})|K^{-1}
|\langle\Phi|\tau+t',l\rangle|
\stackrel{l\to\infty}{\longrightarrow}\infty,
\een }
since the first factor diverges and the second remains bounded and does
not tend to zero. If follows
 $\Phi \notin D(V(\gamma,z)^*)$, hence $V(\gamma,z)$ is noncloseable.\\
It remains to prove the noncloseability of $V(\gamma,z)_0$. The  states
\ben\label{7}
|t,\beta,l,M \rangle := N_l \sum^M_{\alpha_1, \ldots
,\alpha_l=0}\prod^l_{i=1} \frac{(\beta t_i)^{\alpha_i}}{(\alpha_i !
i^{\alpha_i})^{1/2}} \Phi_{\alpha}
\een
are in $D(V(\gamma,z)_0)$, and surely
\ben\label{8}
|t,\beta,l,M\rangle \longrightarrow |t,\beta,l \rangle  \quad (M \to
\infty).
\een
Again by Theorem 2, \cite{BC} it follows from the existence of the
(matrix--)product (Lemma 3, (ii)) that for any $l$
\[
V( \gamma,z) |t,\beta,l,M \rangle \longrightarrow V(\gamma,z)|t,\beta,l
\rangle \quad (M \to \infty).
\]
Hence we can choose a sequence $M_l\to \infty \: (l \to \infty)$ for which
\[
|\langle \Phi|V(\gamma,z)|t,\beta,l,M_l\rangle| \longrightarrow \infty
\quad (l \to \infty).
\]
\hfill$\Box$
\section{Extension to WZNW-models}
The Fock--spaces used for WZNW--models over the affine Lie algebra
$\hat{\frg}$ have the form ${\cal H}_1\otimes \cdots \otimes{\cal H}_l
\otimes {\cal H}_1'\otimes \cdots \otimes {\cal H}_p'$, where $l$ is the
rank of $\frg$ and $p$ the number of positive roots.\\
In ${\cal H}_i$ act free bosonic fields $\beta_i(z), \gamma_i(z)$
$(i=1,\ldots,l)$ and in ${\cal H}_j'$ act free scalar fields $\Phi_j(z)$,
$(j=1,\ldots,p)$. The vertex operators constructed in WZNW--models are
given by expressions of the  form (see for example (3.12) in
\cite{BMP})\\[5mm]
%{\small\mathindent0mm\[
$P_1\left(\beta_1(z),\gamma_1(z)\right)\otimes \cdots \otimes
P_l\left(\beta_l(z),\gamma_l(z)\right) \otimes :
\exp\left(-i\alpha_1\Phi_1(z)\right): \otimes \cdots
\otimes :\exp \left(-i\alpha_p \Phi_p(z)\right):$\\[5mm]
%\]}
were $P_i$ are polynominals and $:\quad:$ denotes Wick--ordering.\\
Of course, our proof applies to $:\exp\left(-i\alpha_i \Phi_i(z)\right):$
as operator in ${\cal H}_i'$. Since the tensor product of an uncloseable
operator with another operator is again uncloseable, Theorem 1 holds in
these cases too.
\section{Appendix: Proof of Lemma 1}
We shall prove the equivalent assertion, that if $t \in l_{2,h}$ and
$\langle \Phi|t+t'\rangle=0$ for all $t' \in l_{2,0}$, then $\Phi=0$.

Let $t \in l_{2,h}$ be fixed. Choose now $r \in \NN$, $\mu_1, \ldots ,
\mu_r \in \NN$. Then for any $u_1, \ldots ,u_r \in \CC$ we can choose
$t^{(l)}\in l_{2,0}$, such that $t+t^{(l)}= (0, \ldots ,
u_1,0,\ldots,u_r,0,0,\ldots,t_l,t_{l+1},\ldots)$. Hence
$t+t^{(l)} \longrightarrow (0,\ldots,u_1,\ldots,u_r,0,0,\ldots)$ for
$l\to\infty$. Since the mapping $t\mapsto |t\rangle$ from $l_{2,h}$ to
$\cal H$ is continous \cite{KS}, we get
\ben
\langle \Phi| t+t^{(l)} \rangle \longrightarrow \langle \Phi|u\rangle
\quad (l\to\infty)
\een
and by standard manipulations for coherent states with $\Phi=\sum
c_{\alpha} \Phi_{\alpha}$
\ben\label{10}
\langle
\Phi|u\rangle=\sum_{\alpha_{\mu_1},\ldots,\alpha_{\mu_r}=0}^{\infty}
\bar{c}_{\alpha} \prod_{i=1}^r \frac{u_i^{\alpha_{\mu_i}}}{\alpha_{\mu_i!}
\mu_i^{\alpha_{mu_i}}}.
\een
which is an entire function in the variables $u_1,\ldots, u_r$. If
$\langle \Phi|t+t^{(l)}\rangle =0$ for all $l$, we get $\langle
\Phi|u\rangle=0$. Since $u_1,\ldots,u_r \in \CC$ are arbitrary the entire
function vanishes identically, and we can conclude $c_{\alpha}=0$ for all
$\alpha$ appearing in the right hand side of (\ref{10}), i.e. for all
$\alpha$ which are of the form
\[
\alpha=(0,\ldots,\alpha_{\mu_1},0,\ldots,\alpha_{\mu_r},0,0,\ldots).
\]
Since $r$ and $\mu_1,\ldots,\mu_r$ are arbitrary, we have proved
$c_{\alpha}=0$ for all $\alpha$, and hence we have $\Phi=0$ as claimed.
\hspace*{\fill}$\Box$\\

\end{document}